\documentclass[twocolumn,aps,prl,superscriptaddress]{revtex4}

\usepackage{sidecap}
\usepackage{ulem}
\usepackage{epsfig}
\usepackage{amsmath,amssymb,amsthm}
\usepackage{graphicx}
\usepackage{bm}
\usepackage{color,soul}

\DeclareMathAlphabet{\mathpzc}{OT1}{pzc}{m}{it}

\begin{document}

\renewcommand{\textfraction}{0.00}


\newcommand{\vAi}{{\cal A}_{i_1\cdots i_n}}
\newcommand{\vAim}{{\cal A}_{i_1\cdots i_{n-1}}}
\newcommand{\vAbi}{\bar{\cal A}^{i_1\cdots i_n}}
\newcommand{\vAbim}{\bar{\cal A}^{i_1\cdots i_{n-1}}}
\newcommand{\htS}{\hat{S}}
\newcommand{\htR}{\hat{R}}
\newcommand{\htB}{\hat{B}}
\newcommand{\htD}{\hat{D}}
\newcommand{\htV}{\hat{V}}
\newcommand{\cT}{{\cal T}}
\newcommand{\cM}{{\cal M}}
\newcommand{\cMs}{{\cal M}^*}
\newcommand{\vk}{\vec{\mathbf{k}}}
\newcommand{\bk}{\bm{k}}
\newcommand{\kt}{\bm{k}_\perp}
\newcommand{\kp}{k_\perp}
\newcommand{\km}{k_\mathrm{max}}
\newcommand{\vl}{\vec{\mathbf{l}}}
\newcommand{\bl}{\bm{l}}
\newcommand{\bK}{\bm{K}}
\newcommand{\bb}{\bm{b}}
\newcommand{\qm}{q_\mathrm{max}}
\newcommand{\vp}{\vec{\mathbf{p}}}
\newcommand{\bp}{\bm{p}}
\newcommand{\vq}{\vec{\mathbf{q}}}
\newcommand{\bq}{\bm{q}}
\newcommand{\qt}{\bm{q}_\perp}
\newcommand{\qp}{q_\perp}
\newcommand{\bQ}{\bm{Q}}
\newcommand{\vx}{\vec{\mathbf{x}}}
\newcommand{\bx}{\bm{x}}
\newcommand{\tr}{{{\rm Tr\,}}}
\newcommand{\bc}{\textcolor{blue}}

\newcommand{\beq}{\begin{equation}}
\newcommand{\eeq}[1]{\label{#1} \end{equation}}
\newcommand{\ee}{\end{equation}}
\newcommand{\bea}{\begin{eqnarray}}
\newcommand{\eea}{\end{eqnarray}}
\newcommand{\beqar}{\begin{eqnarray}}
\newcommand{\eeqar}[1]{\label{#1}\end{eqnarray}}

\newcommand{\half}{{\textstyle\frac{1}{2}}}
\newcommand{\ben}{\begin{enumerate}}
\newcommand{\een}{\end{enumerate}}
\newcommand{\bit}{\begin{itemize}}
\newcommand{\eit}{\end{itemize}}
\newcommand{\ec}{\end{center}}
\newcommand{\bra}[1]{\langle {#1}|}
\newcommand{\ket}[1]{|{#1}\rangle}
\newcommand{\norm}[2]{\langle{#1}|{#2}\rangle}
\newcommand{\brac}[3]{\langle{#1}|{#2}|{#3}\rangle}
\newcommand{\hilb}{{\cal H}}
\newcommand{\pleft}{\stackrel{\leftarrow}{\partial}}
\newcommand{\pright}{\stackrel{\rightarrow}{\partial}}

\title{Joint $R_{AA}$ and $v_2$ predictions for $Pb+Pb$ collisions at the LHC within DREENA-C framework}

\author{Dusan Zigic}
\affiliation{Institute of Physics Belgrade, University of Belgrade, Serbia}

\author{Igor Salom}
\affiliation{Institute of Physics Belgrade, University of Belgrade, Serbia}

\author{Jussi Auvinen}
\affiliation{Institute of Physics Belgrade, University of Belgrade, Serbia}

\author{Marko Djordjevic}
\affiliation{Faculty of Biology, University of Belgrade, Serbia}

\author{Magdalena Djordjevic\footnote{E-mail: magda@ipb.ac.rs}}
\affiliation{Institute of Physics Belgrade, University of Belgrade, Serbia}

\begin{abstract} In this paper, we presented our recently developed DREENA-C framework, which is a fully optimized computational suppression procedure based on our state-of-the-art dynamical energy loss formalism in constant temperature finite size QCD medium. With this framework, we for the first time, generated joint $R_{AA}$ and $v_2$ predictions within our dynamical energy loss formalism. The predictions are generated for both light and heavy flavor probes, and different centrality regions in $Pb+Pb$ collisions at the LHC, and  compared with the available experimental data. Despite the fact that DREENA-C does not contain medium evolution (to which $v_2$ is largely sensitive) and the fact that other approaches faced difficulties in explaining $v_2$ data, we find that DREENA-C leads to qualitatively good agreement with this data, though quantitatively, the predictions are visibly above the experimental data. Intuitive explanation behind such results is presented, supporting the validity of our model, and it is expected that introduction of evolution in the ongoing improvement of DREENA framework, will lead to better joint agreement with $R_{AA}$ and $v_2$ data, and allow better understanding of the underlying QCD medium.
\end{abstract}

\maketitle

\section{Introduction}

Quark-gluon plasma (QGP) is a new state of matter~\cite{Collins,Baym} consisting of interacting quarks, antiquarks and gluons. Such new state of matter is created in ultra-relativistic heavy ion collisions at Relativistic Heavy Ion Collider (RHIC) and Large Hadron Collider (LHC). Rare high momentum probes, which are created in such collisions and which transverse QGP, are excellent probes of this extreme form of matter~\cite{QGP1,QGP2,QGP3}. Different observables (such as angular averaged nuclear modification factor $R_{AA}$ and angular anisotropy $v_2$), together with probes with different masses, probe this medium in a different manner. Therefore, comparing comprehensive set of joint predictions for different probes and observables, with available experimental data at different experiments, collision systems and collision energies, allows investigating properties of QCD medium created in these collisions~\cite{Bjorken,DG_PRL,Kharzeev}, i.e. QGP tomography.

With this goal in mind, we developed state-of-the-art dynamical energy loss formalism~\cite{MD_PRC,DH_PRL}, which includes different important effects. Namely, {\it i)} contrary to the widely used approximation of static scattering centers, this formalism takes into account that QGP consists of dynamical (that is moving) partons and that created medium has finite size. {\it ii)} The calculations are based on the finite temperature field theory~\cite{Kapusta,Le_Bellac}, and generalized HTL approach. {\it iii)} The formalism takes into account both radiative~\cite{MD_PRC} and collisional~\cite{MD_Coll} energy losses, is applicable to both light and heavy flavor, and has been generalized to the case of finite magnetic~\cite{MD_MagnMass} mass and running coupling~\cite{MD_PLB}.
This formalism was further integrated into numerical procedure~\cite{MD_PLB}, which includes initial $p_\perp$ distribution of leading partons~\cite{Vitev0912}, energy loss with path-length~\cite{Dainese,WHDG} and multi-gluon~\cite{GLV_suppress} fluctuations, and fragmentation functions~\cite{DSS,BCFY,KLP}, to generate the final medium modified distribution of high pt hadrons. 

However, due to the complexity of this model, the model does not take into account the medium evolution, which is a necessary ingredient for QGP tomography. Therefore, our future major task is redeveloping the dynamical energy loss model, and the corresponding numerical procedure, so that it accounts for evolving QGP medium. Equally important, the framework has to be able to efficiently generate predictions to be compared with a wide range of available (or upcoming) experimental data. Within this, all ingredients stated above have to be kept (with no additional simplifications used in the numerical procedure), as all of these ingredients were shown to be important for reliable theoretical predictions of jet suppression~\cite{BD_JPG}.

As a first step towards this major goal, we here developed a numerical framework DREENA-C (Dynamical Radiative and Elastic ENergy loss Approach), where "C" denotes constant temperature QCD medium. This framework is in its essence equivalent to the numerical procedure presented in~\cite{MD_PLB}, with the difference that the code is now optimized to use minimal computer resources and produce predictions within more than two orders of magnitude shorter time compared to~\cite{MD_PLB}. Such step is necessary, as all further improvements of the framework, necessarily need significantly more computer time and resources, so without this step, further improvements would not be realistically possible. That is, DREENA-C framework, addresses the goal of efficiently generating predictions for diverse observables, while the goal of introducing the medium evolution will be a subject of our future work.

With DREENA-C framework, we will in this paper, for the first time, present joint $R_{AA}$ and $v_2$ theoretical predictions within our dynamcial energy loss formalism. These predictions will be generated for all available light and heavy observables in $Pb+Pb$ collisions at the LHC, and for various centrality regions. Motivation for generating these predictions is the following: {\it i)} The theoretical models up-to-now were not able to jointly explain these data, which is known as v2 puzzle~\cite{v2Puzzle}. That is, the models lead to underprediction of $v_2$, unless new phenomena (e.g. magnetic monopoles) are introduced~\cite{CUJET3}. {\it ii)} Having this puzzle in mind, comparison of our theoretical predictions with comprehensive set of experimental $R_{AA}$ and $v_2$ data, allows testing to what extent state-of-the-art energy loss model, but with no QGP evolution included, is able to jointly explain these data. {\it iii)} The predictions will establish an important baseline for testing how future introduction of the medium evolution will improve the formalism.

\section{Methods}

DREENA-C framework is fully optimized numerical procedure, which contains all ingredients presented in detail in~\cite{MD_PLB}. We below briefly outline the main steps in this procedure.

The  quenched spectra of light and heavy flavor observables are calculated according to the generic pQCD convolution:
\begin{eqnarray}
\frac{E_f d^3\sigma}{dp_f^3} &=& \frac{E_i d^3\sigma(Q)}{dp^3_i}
 \otimes
{P(E_i \rightarrow E_f )}
\nonumber \\
&& \otimes D(Q \to H_Q)  \otimes f(H_Q \to e, J/\psi). \;
\label{schem} \end{eqnarray}

Subscripts "$i$"  and "$f$" correspond, respectively, to "initial" and "final", and $Q$ denotes initial light or heavy flavor jet. $E_i d^3\sigma(Q)/dp_i^3$ denotes the initial momentum spectrum, which are computed according to~\cite{Vitev0912},  $P(E_i \rightarrow E_f )$ is the energy loss probability, computed within the dynamical energy loss formalism~\cite{MD_PRC,DH_PRL}, with multi-gluon~\cite{GLV_suppress}, path-length fluctuations~\cite{WHDG} and running coupling~\cite{MD_PLB}. $D(Q \to H_Q)$ is the fragmentation function of light and heavy flavor parton $Q$ to hadron $H_Q$, where for light flavor, D and B mesons we use, DSS~\cite{DSS}, BCFY~\cite{BCFY} and KLP~\cite{KLP} fragmentation functions, respectively.
\medskip

As noted above, we model the medium by assuming a constant average temperature of QGP. We concentrate on $5.02$~TeV $Pb+Pb$ collisions at the LHC, though we note that these predictions will be applicable for $2.76$~TeV $Pb+Pb$ collisions as well, since the predictions for these two collision energies almost overlap~\cite{MD_5TeV}. The temperatures for different centralities in $5.02$~TeV $Pb+Pb$ collisions are calculated according to~\cite{DDB_PLB}. As a starting point in this calculation we use the effective temperature ($T_{eff}$) of 304 MeV for 0-40$\%$ centrality $2.76$ TeV Pb+Pb collisions at the LHC~\cite{ALICE_T} experiments (as extracted by ALICE); e.g. leading to 335 MeV temperature for 0-10\% central $5.02$ TeV collisions (for more details see~\cite{MD_5TeV}). Path-length distributions are calculated following the procedure described in~\cite{Dainese}, with an additional hard sphere restriction $r < R_A$ in the Woods-Saxon nuclear density distribution to regulate the path lengths in the peripheral collisions.

In numerical calculations, we consider a QGP  with $\Lambda_{QCD}=0.2$~GeV and $n_f{\,=\,}3$. The temperature dependent Debye mass $\mu_E (T)$  is obtained from~\cite{Peshier}, while for the light quarks, we assume that their mass is dominated by the thermal mass $M{\,\approx\,}\mu_E/\sqrt{6}$, and the gluon mass is $m_g\approx \mu_E/\sqrt{2}$~\cite{DG_TM}. The charm (bottom) mass is $M{\,=\,}1.2$\,GeV ($M{\,=\,}4.75$\,GeV). Finite magnetic mass effect is also included in our framework~\cite{MD_MagnMass}, as various non-perturbative calculations~\cite{Maezawa,Nakamura,Hart,Bak} have shown that magnetic mass $\mu_M$ is different from zero in QCD matter created at the LHC and RHIC. Magnetic to electric mass ratio is extracted from these calculations to be $0.4 < \mu_M/\mu_E < 0.6$, so presented uncertainty in the predictions, will come from this range of screening masses ratio. Note that we use no fitting parameters in comparison with the data, i.e. all the parameters correspond to standard literature values.

\section{Results and discussion}

\begin{figure*}
\epsfig{file=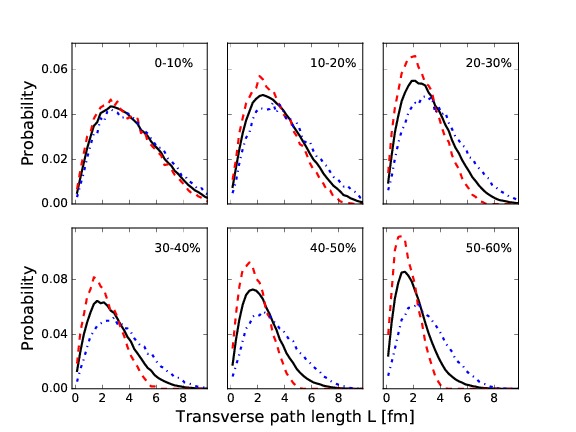,width=5in,height=3.5in,clip=5,angle=0}
\vspace*{-0.4cm}
\caption{{\bf Path-length distributions.} Probability distributions for hard parton path lengths in $Pb+Pb$ collisions at $\sqrt{s_{NN}}=5.02$ TeV for (0-10)\% - (50-60)\% centrality classes. Solid black curves: the total distributions with all hard partons included are represented; Dashed red curves: the distributions include only in-plane particles ($|\phi|<15^{\circ}$ or $||\phi|-180^{\circ}|<15^{\circ}$); Dash-dotted blue curves: the distributions include only out-of-plane partons ($||\phi|-90^{\circ}<15^{\circ}$).}
\label{PathLength}
\end{figure*}

In this section, we will generate joint $R_{AA}$ and $v_2$ predictions for charged hadrons, D and B mesons in $Pb+Pb$ collisions at the LHC. In Figure~\ref{PathLength} we first show probability distributions for hard parton path lengths in Pb+Pb collisions for different centralities, obtained by the procedure specified in the previous section. For most central collisions, we observe that in-plane and out-of-plane distributions almost overlap with the total (average) path-length distributions, as expected. As the centrality increases, in-plane and out-of-plane distributions start to significantly separate (in different directions) from average path-length distributions. Having in mind that
\begin{eqnarray}
v_2 \approx \frac{1}{2} \frac{R_{AA}^{in}-R_{AA}^{out}}{R_{AA}^{in}+R_{AA}^{out}},
\label{v2} \end{eqnarray}
this leads the expectation of $v_2$ being small in most central collisions and increasing with increasing centrality.
\begin{figure}
\epsfig{file=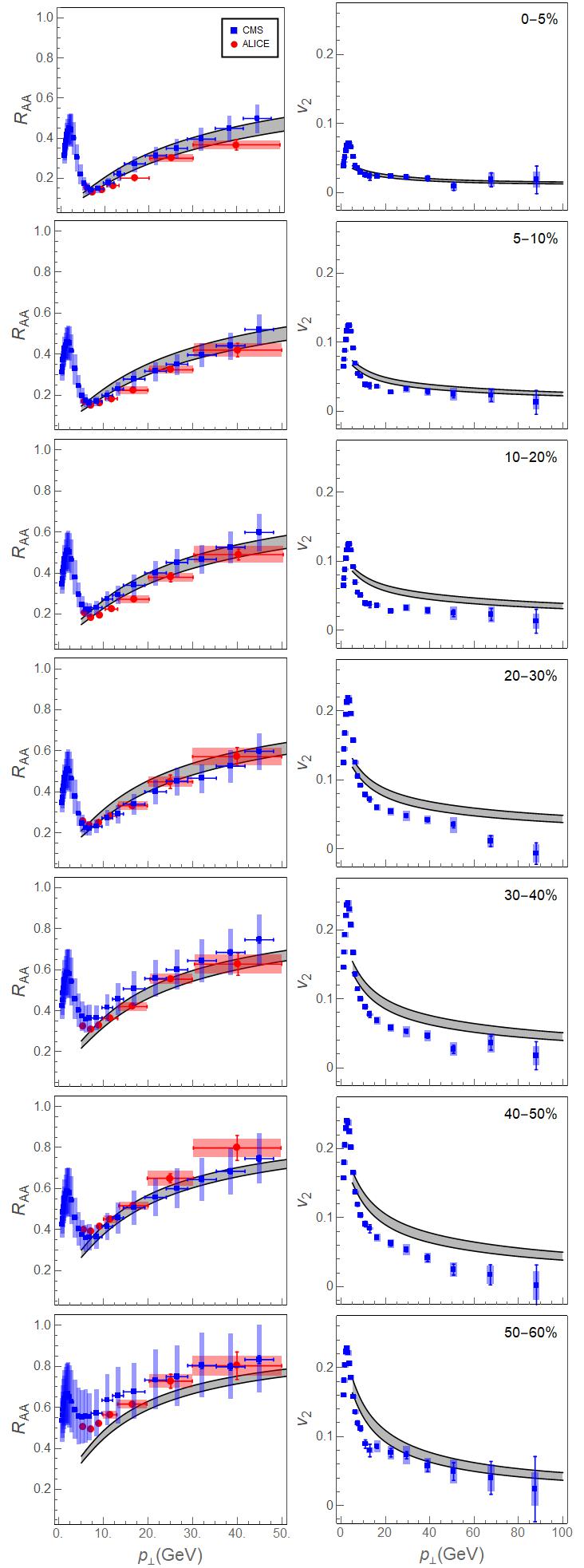,width=3.2in,height=8in,clip=5,angle=0}
\vspace*{-0.2cm}
\caption{ {\bf Joint $R_{AA}$ and $v_2$ predictions for charged hadrons.}{\it Left panels:} Theoretical predictions for $R_{AA}$ {\it vs.} $p_\perp$ are compared with ALICE~\cite{ALICE_CH_RAA} (red circles) and CMS~\cite{CMS_CH_RAA} (blue squares) charged hadron experimental data for $5.02$~TeV $Pb+Pb$ collisions at the LHC.  {\it Right panels:} Theoretical predictions for $v_2$ {\it vs.} $p_\perp$ are compared with CMS~\cite{CMS_CH_v2} (blue squares) charged hadron experimental data for $5.02$~TeV $Pb+Pb$ collisions at the LHC. On each panel, the upper (lower) boundary of each gray band corresponds to $\mu_M/\mu_E =0.6$ ($\mu_M/\mu_E =0.4$). Rows 1-7 correspond to, respectively, $0-5\%$, $5-10\%$, $10-20\%$,..., $50-60\%$ centrality regions. }
\label{CH_RAA_v2}
\end{figure}

Based on path-length distributions from Figure~\ref{PathLength}, we can now calculate average $R_{AA}$, as well as in-plane and out-of-plane $R_{AA}$s ($R_{AA}^{in}$ and $R_{AA}^{out}$), and consequently $v_2$ for both light and heavy flavor probes and different centralities. We start by generating predictions for charged hadrons, where data for both $R_{AA}$ and $v_2$ are available. Comparison of our joint predictions with experimental data is shown in Figure~\ref{CH_RAA_v2}, where left and right panels correspond, respectively, to $R_{AA}$ and $v_2$. We see good agreement with $R_{AA}$ data, which is expected based on our previous studies~\cite{MD_PRL,MD_PLB,DDB_PLB,MD_5TeV,DBZ,MD_PLB16}. Regarding $v_2$, we surprisingly see that our model actually leads to qualitatively good agreement with the data.  Even more surprisingly, we see that our $v_2$ predictions are visibly above the data. This is in contrast with other energy loss models which consistently lead to underprediction of $v_2$, where, to resolve this, new phenomena (e.g. magnetic monopoles) were introduced~\cite{CUJET3}.

\begin{figure*}
\epsfig{file=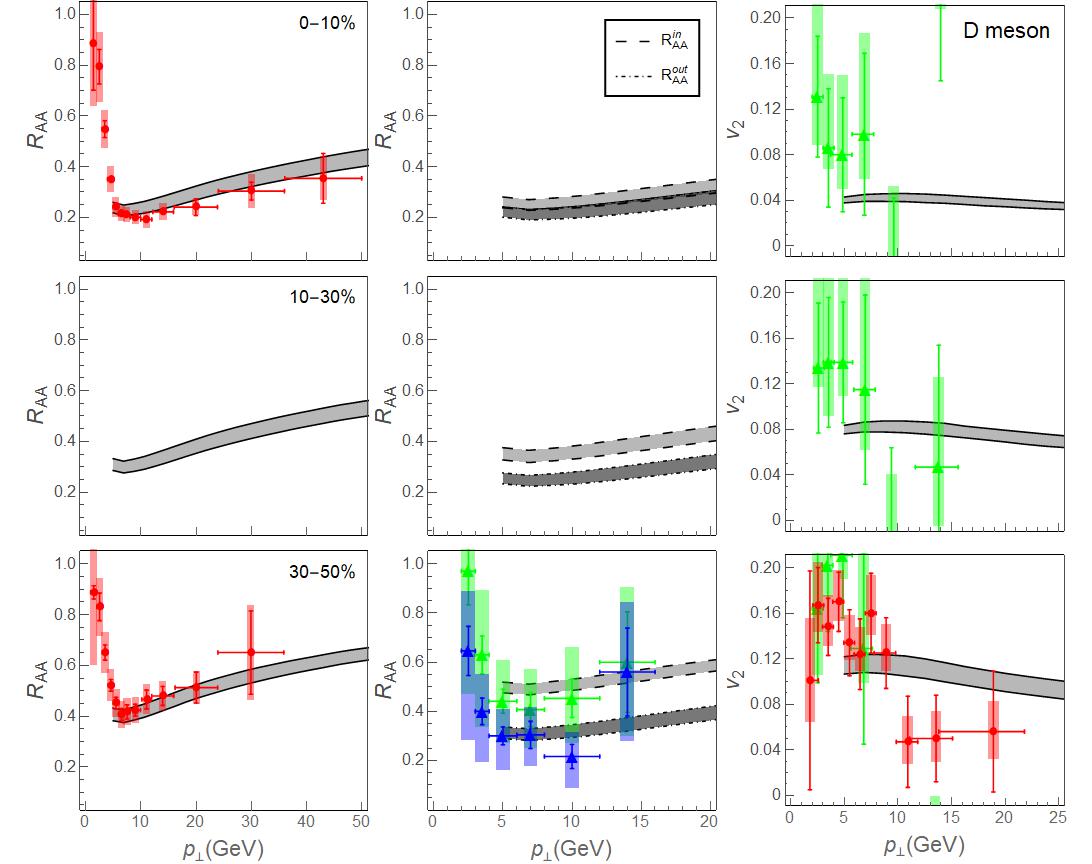,width=4.8in,height=3.5in,clip=5,angle=0}
\vspace*{-0.2cm}
\caption{{\bf Joint $R_{AA}$, $R_{AA}^{in}$, $R_{AA}^{out}$ and $v_2$ predictions for D mesons.} {\it Left panels:} Theoretical predictions for $R_{AA}$ {\it vs.} $p_\perp$ are compared with ALICE~\cite{ALICE_D_RAA} (red circles) D meson experimental data for $5.02$~TeV $Pb+Pb$ collisions at the LHC. {\it Middle panels:} Theoretical predictions for in-plane (dashed curves) and out-of-plane (dot-dashed curves) $R_{AA}$s {\it vs.} $p_\perp$ are compared with ALICE~\cite{ALICE_D_v2} D meson experimental data (green and blue triangles, respectively) for $2.76$~TeV $Pb+Pb$ collisions at the LHC. {\it Right panels:} Theoretical predictions for $v_2$ {\it vs.} $p_\perp$ are compared with ALICE~\cite{ALICE_D_v2_5} D meson experimental data for $5.02$~TeV (red circles) and $2.76$~TeV (green triangles) $Pb+Pb$ collisions at the LHC. On each panel, the upper (lower) boundary of each gray band corresponds to $\mu_M/\mu_E =0.6$ ($\mu_M/\mu_E =0.4$). First to third row correspond to, respectively, $0-10\%$, $10-30\%$ and $30-50\%$ centrality regions. }
\label{D_RAA_V2}
\end{figure*}

In Figure~\ref{D_RAA_V2}, we provide predictions for D meson average $R_{AA}$ (left panel), in-plane and out-of-plane $R_{AA}$ (middle panel) as well as $v_2$ (right panel) for three different centrality regions. Predictions are compared with available experimental data, where data for both $5.02$~TeV and $2.76$~TeV are shown (since the data for $5.02$~TeV are still scarce, and the data for these two collision energies overlap with eachother~\cite{MD_5TeV}). For average, in-plane and out-of-plane $R_{AA}$, we observe good agreement with the data (where available). Agreement with $v_2$ data is also qualitatively good (with predictions again above the experimental data), though we note that the error bars are large.

\begin{figure*}
\epsfig{file=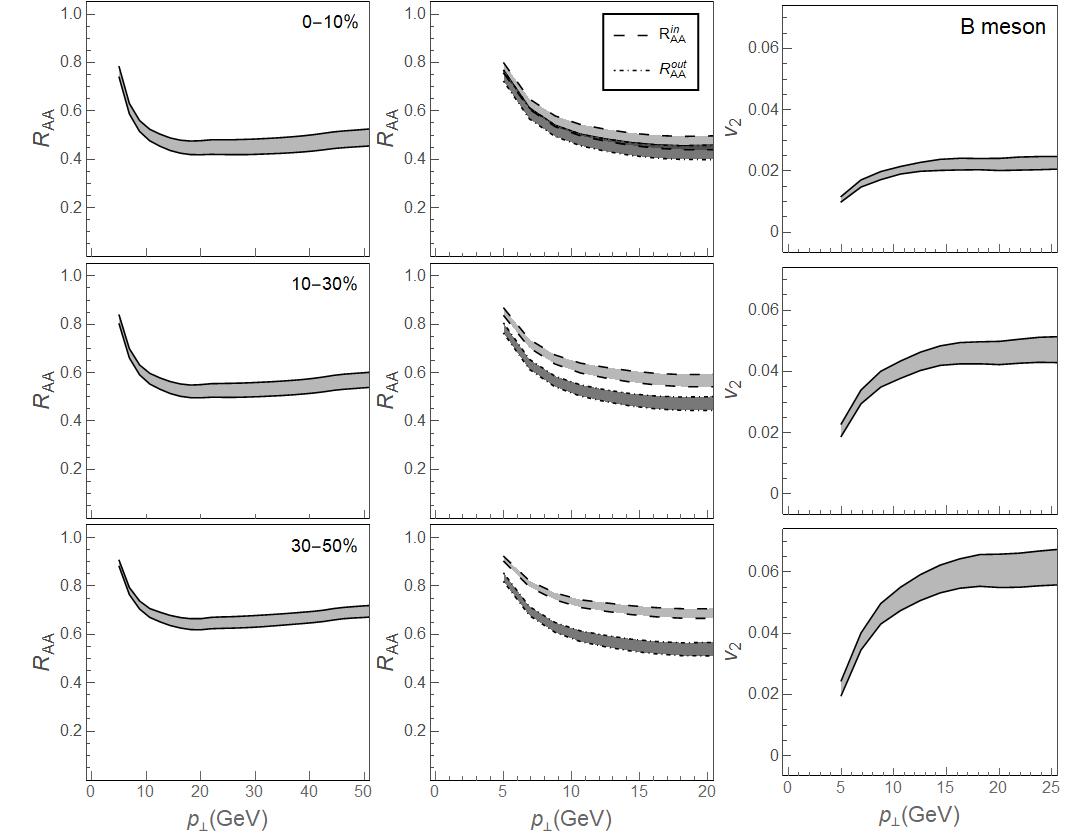,width=4.8in,height=3.5in,clip=5,angle=0}
\vspace*{-0.2cm}
\caption{{\bf Joint $R_{AA}$, $R_{AA}^{in}$, $R_{AA}^{out}$ and $v_2$ predictions for B mesons.} {\it Left panels:} Theoretical predictions for B meson $R_{AA}$ {\it vs.} $p_\perp$ are shown. {\it Middle panels:} Theoretical predictions for B-meson in-plane (dashed curves) and out-of-plane (dot-dashed curves) $R_{AA}$s {\it vs.} $p_\perp$ are shown. {\it Right panels:} Theoretical predictions for B meson $v_2$ {\it vs.} $p_\perp$ are shown. On each panel, the upper (lower) boundary of each gray band corresponds to $\mu_M/\mu_E =0.6$ ($\mu_M/\mu_E =0.4$). First to third row correspond to, respectively, $0-10\%$, $10-30\%$ and $30-50\%$ centrality regions. }
\label{B_RAA_V2}
\end{figure*}
Figure~\ref{B_RAA_V2} shows equivalent predictions as Figure~\ref{D_RAA_V2}, only for B mesons. While B meson experimental data are yet to become available, we predict $v_2$ which is significantly different from zero for all centrality regions. This does not necessarily mean that heavy B meson flows, as flow is inherently connected with {\it low} $p_\perp$ $v_2$, and here we show predictions for high $p_\perp$. On the other hand, high $p_\perp$ $v_2$ is connected with the difference in the energy loss (i.e. suppression) for particles going in different (e.g. in-plane and out-of-plane) directions. From the middle panels, we see significant difference between in-plane and out-of-plane $R_{AA}$, which is a consequence of the difference in the path-lengths shown in Fig.~\ref{PathLength}. This difference then leads to our predictions of non zero $v_2$ for {\it high} $p_\perp$ B mesons.

Overall, while we see that our predicted $R_{AA}$s agree well with all measured (light and heavy flavor) data, our $v_2$ predictions are consistently above the experimental data. This observation then leads to the following two questions: {\it i)} What is the reason behind the observed overestimation of $v_2$ within DREENA-C framework, and can expanding medium lead to a better agreement with the experimental data? {\it ii)} Do we expect that B meson $v_2$ predictions will still be non-zero, once the expanding medium is introduced?

To intuitively approach these questions, note that, within our dynamical energy loss formalism, $\Delta E/E \sim T^a$ and $\Delta E/E \sim L^b$, where $a, b \rightarrow 1$ ($\Delta E/E$ is fractional energy loss, $T$ is the average temperature of the medium, while $L$ is the average path-length traversed by the jet). To be more precise, note that both dependencies are close to linear, though still significantly different from 1~\cite{MD_5TeV}). However, for the purpose of this estimate, let us assume that both $a$ and $b$ are equal to 1, leading to
\begin{eqnarray}
\Delta E/E \approx \eta T L,
\label{ElossEstimate} \end{eqnarray}
where $\eta$ is a proportionality factor.

Another, commonly used estimate~\cite{GLV_suppress} is that
\begin{eqnarray}
R_{AA} \approx (1-\frac{1}{2} \frac{\Delta E}{E})^{n-2},
\label{RaaEstimate} \end{eqnarray}
where $n$ is the steepness of the initial momentum distribution function.

For small fractional energy loss, Eq.~\ref{RaaEstimate} becomes
\begin{eqnarray}
R_{AA} \approx (1-\frac{n-2}{2} \frac{\Delta E}{E}) \approx (1-\xi T L),
\label{RaaEstimate2} \end{eqnarray}
where $\xi = (n-2) \eta/2$.

\smallskip

In DREENA-C approach, T is constant, and the same in in-plane and out-of-plane directions, while $L_{in}=L-\Delta L$ and $L_{out}=L+\Delta L$, leading to
\begin{eqnarray}
R_{AA} &\approx& \frac{1}{2} (R_{AA}^{in} + R_{AA}^{out}) \approx \frac{1}{2} (1-\xi T L_{in} + 1-\xi T L_{out}) \nonumber \\
&=& 1-\xi T \frac{L_{in}+L_{out}}{2} = 1-\xi T L,
\label{RaaAver} \end{eqnarray}
and
\begin{eqnarray}
v_2 &\approx& \frac{1}{2} \frac{R_{AA}^{in}-R_{AA}^{out}}{R_{AA}^{in}+R_{AA}^{out}} \approx \frac{1}{2} \frac{(1-\xi T L_{in}) - (1-\xi T L_{out})}{1-\xi T L} \nonumber \\
&=& \frac{1}{2} \frac{\xi T \Delta L}{1-\xi T L} \approx \frac{ \xi T \Delta L}{2}.
\label{v2Aver} \end{eqnarray}

If the medium evolves, the average temperature along in-plane will be larger than along out-of-plane direction, leading to $T_{in}=T+\Delta T$ and $T_{out}=T-\Delta T$. By repeating the above procedure in this case, it is straightforward to obtain
\begin{eqnarray}
R_{AA} &\approx&  1-\xi T L
\label{RaaEvol} \end{eqnarray}
and
\begin{eqnarray}
v_2 &\approx& \frac{1}{2} \frac{(1-\xi T_{in} L_{in}) - (1-\xi T_{out} L_{out})}{1-\xi T L} \nonumber \\
&=& \frac{1}{2} \frac{\xi T \Delta L - \xi \Delta T L }{1-\xi T L} \approx \frac{\xi T \Delta L - \xi \Delta T L}{2}.
\label{v2Evol} \end{eqnarray}

Therefore, from the above estimates, we see that $R_{AA}$ should not be very sensitive to the medium evolution, while $v_2$ will be quite sensitive to this evolution, as previously noted elsewhere~\cite{Molnar,Thorsten}. Moreover, from Eqs.~(\ref{v2Aver}) and~(\ref{v2Evol}), we see that introduction of temperature evolution is expected to lower $v_2$ compared to constant $T$ case. Consequently, accurate energy loss models applied to non-evolving medium should lead to higher $v_2$ than expected, and introduction of $T$ evolution in such models would lower the $v_2$ compared to non-evolving case. Based on this, and the fact that previous theoretical approaches were not able to reach high enough $v_2$ without introducing new phenomena~\cite{CUJET3}, we argue that the dynamical energy loss formalism has the right features needed to accurately describe jet-medium interactions in QGP. 

Regarding the second question mentioned above, for B meson to have $v_2 \approx 0$, it is straightforward to see that one needs $\Delta T/T \approx \Delta L/L$. Having in mind that $\Delta L/L$ is quite large for larger centralities (see Fig.~\ref{PathLength}), $\Delta T/T$ would also have to be about the same magnitude. We do not expect this to happen, based on our preliminary estimates of the temperature changes in in-plane and out-of-plane in Bjorken expansion scheme. That is, our expectations is that B meson $v_2$ will be smaller than presented here, but still significantly larger than zero, at least for large centrality regions. However, this still remains to be tested in the future with the introduction of full evolution model within our framework.

\section{Conclusion}
In this paper, we introduced our recently developed DREENA-C framework, which is computational suppression procedure based on our dynamical energy loss formalism in constant $T$ finite size QCD medium. The framework is conceptually equivalent to the numerical procedure developed in ~\cite{MD_PLB}, and consequently lead to the same numerical results. However, from practical perspective, DREENA-C is computationally fully optimized and more than two orders of magnitude faster than its predecessor.

We here used DREENA-C framework to, for the first time, generate joint $R_{AA}$ and $v_2$ predictions for both light and heavy flavor probes and different centrality regions in $Pb+Pb$ collisions at the LHC, and compare them with the available experimental data. Having in mind that DREENA-C does not contain medium evolution, and that $v_2$ is largely sensitive to the QGP evolution, we did not expect that our model would lead to good agreement with the data. However, contrary to these expectations, and to the fact that other approaches faced difficulties in jointly explaining $R_{AA}$ and $v_2$ data, we actually find that DREENA-C leads to good agreement with $R_{AA}$ data and qualitatively good agreement with $v_2$ data (though quantitatively, the predictions overestimate the data). Intuitive explanation behind such results is presented, supporting the validity of our model, with an expectation that introduction of evolution in the DREENA framework will improve the agreement with $v_2$ data. These results therefore further confirm that our dynamical energy loss formalism is a suitable basis for the QGP tomography, which is the main goal for our future research.

{\em Acknowledgments:}
This work is supported by the European Research Council, grant ERC-2016-COG: 725741, and by  and by the Ministry of Science and Technological
Development of the Republic of Serbia, under project numbers ON171004 and ON173052.


\begin{references}

\bibitem{Collins} J. C. Collins and M. J. Perry, Phys. Rev. Lett. {\bf 34}, 1353 (1975).

\bibitem{Baym} G. Baym and S. A. Chin, Phys. Lett. B {\bf 62}, 241 (1976).

\bibitem{QGP1} M. Gyulassy and L. McLerran, Nucl. Phys. A {\bf 750}, 30 (2005).

\bibitem{QGP2} E. V. Shuryak, Nucl. Phys. A {\bf 750}, 64 (2005).

\bibitem{QGP3} B. Jacak and P. Steinberg, Phys. Today {\bf 63}, 39 (2010).

\bibitem{Bjorken} J. D. Bjorken: FERMILAB-PUB-82-059-THY (1982) 287, 292

\bibitem{DG_PRL} M. Djordjevic, M. Gyulassy and S. Wicks, Phys. Rev. Lett. {\bf 94}, 112301 (2005).

\bibitem{Kharzeev} Yu. L. Dokshitzer and D. Kharzeev, Phys. Lett. B {\bf 519}, 199 (2001).

\bibitem{MD_PRC}  M. Djordjevic, Phys. Rev. C {\bf 80}, 064909 (2009).

\bibitem{DH_PRL}  M. Djordjevic and U. Heinz, Phys. Rev. Lett. {\bf 101},
022302 (2008).

\bibitem{Kapusta}
  J. I. Kapusta, {\it Finite-Temperature Field Theory} (Cambridge
  University Press, 1989).

\bibitem{Le_Bellac}
  M. Le Bellac, {\it Thermal Field Theory} (Cambridge
  University Press, 1996).

\bibitem{MD_Coll}  M. Djordjevic, Phys. Rev. C {\bf 74}, 064907 (2006).

\bibitem{MD_PLB} M. Djordjevic and M. Djordjevic,  Phys. Lett. B {\bf 734}, 286 (2014).

\bibitem{BD_JPG} B. Blagojevic and M. Djordjevic, J. Phys. G {\bf 42}, 075105 (2015).

\bibitem{Vitev0912} Z. B. Kang, I. Vitev and H. Xing, Phys. Lett. B {\bf 718},
482 (2012), R. Sharma, I. Vitev and B.W. Zhang, Phys. Rev. C {\bf 80},
054902  (2009)

\bibitem{Dainese} A. Dainese, Eur. Phys. J. C {\bf 33}, 495 (2004).

\bibitem{WHDG}  S. Wicks, W. Horowitz, M. Djordjevic and M. Gyulassy, Nucl.
Phys. A {\bf 784}, 426 (2007).

\bibitem{GLV_suppress} M. Gyulassy, P. Levai and I. Vitev, Phys.\ Lett.\ B {\bf 538}, 282 (2002).

\bibitem{DSS} D. de Florian, R. Sassot and M. Stratmann, Phys. Rev. D {\bf 75}, 114010 (2007).

\bibitem{BCFY} M. Cacciari, P. Nason, JHEP {\bf 0309}, 006 (2003),
E. Braaten, K.-M. Cheung, S. Fleming and T. C. Yuan, Phys. Rev. D
{\bf 51}, 4819 (1995)

\bibitem{KLP} V. G. Kartvelishvili, A.K. Likhoded, V.A. Petrov, Phys. Lett. B {\bf 78}, 615 (1978).

\bibitem{v2Puzzle} J.~Noronha-Hostler, B.~Betz, J.~Noronha and M.~Gyulassy, Phys.\ Rev.\ Lett.\  {\bf 116}, no. 25, 252301 (2016)
 
\bibitem{CUJET3}  J. Xu, J. Liao, and M. Gyulassy, Chinese Physics Letters {\bf 32}  092501 (2015);  S.~Shi, J.~Liao and M.~Gyulassy, arXiv:1804.01915 [hep-ph].

\bibitem{Bjorken} J. D. Bjorken, Physical review D {\bf 27}, 140 (1983)

\bibitem{DDB_PLB} M. Djordjevic, M. Djordjevic and B. Blagojevic,  Phys. Lett. B {\bf 737}, 298 (2014).

\bibitem{ALICE_T} M. Wilde (for the ALICE Collaboration) Nucl. Phys. A
{\bf 904-905} 573c (2013)

\bibitem{MD_5TeV} M. Djordjevic and M. Djordjevic, Phys. Rev. C {\bf 92}, 024918 (2015).

\bibitem{Peshier} A. Peshier, hep-ph/0601119 (2006).

\bibitem{DG_TM} M. Djordjevic and M. Gyulassy, Phys.\ Rev.\ C \ {\bf 68}, 034914 (2003).

\bibitem{MD_MagnMass}  M. Djordjevic, Phys. Lett. B {\bf 709}, 229 (2012).

\bibitem{Maezawa} Yu. Maezawa {\it et al.} [WHOT-QCD Collaboration],
Phys. Rev. D {\bf 81} 091501 (2010);

\bibitem{Nakamura} A. Nakamura, T. Saito and S. Sakai, Phys. Rev. D {\bf 69},
014506 (2004).

\bibitem{Hart} A. Hart, M. Laine and O. Philipsen, Nucl. Phys. B {\bf 586},
443 (2000).

\bibitem{Bak} D. Bak, A. Karch, L. G. Yaffe, JHEP {\bf 0708}, 049
(2007).

\bibitem{MD_PRL} M. Djordjevic,  Phys. Rev. Lett. {\bf 734}, 286 (2014).

\bibitem{DBZ} M. Djordjevic, B. Blagojevic and L. Zivkovic, Phys. Rev. C {\bf },  (2016).

\bibitem{MD_PLB16} M. Djordjevic,  Phys. Lett. B {\bf }, (2016).

\bibitem{ALICE_CH_RAA} S.~Acharya {\it et al.} [ALICE Collaboration],
  arXiv:1802.09145 [nucl-ex].

\bibitem{CMS_CH_RAA}  V.~Khachatryan {\it et al.} [CMS Collaboration],  JHEP {\bf 1704}, 039 (2017).

\bibitem{CMS_CH_v2} A.~M.~Sirunyan {\it et al.} [CMS Collaboration], Phys.\ Lett.\ B {\bf 776}, 195 (2018)

\bibitem{ALICE_D_RAA} S.~Jaelani [ALICE Collaboration],  Int.\ J.\ Mod.\ Phys.\ Conf.\ Ser.\  {\bf 46}, 1860018 (2018)

\bibitem{ALICE_D_v2} B.~B.~Abelev {\it et al.} [ALICE Collaboration],  Phys.\ Rev.\ C {\bf 90}, no. 3, 034904 (2014)

\bibitem{ALICE_D_v2_5} S.~Acharya {\it et al.} [ALICE Collaboration],  Phys.\ Rev.\ Lett.\  {\bf 120}, no. 10, 102301 (2018)

\bibitem{Molnar} D. Molnar and D. Sun, Nucl. Phys. A {\bf 932}, 140 (2014); Nucl. Phys. A {\bf 910-911}, 486 (2013).

\bibitem{Thorsten} T. Renk, Phys. Rev. C {\bf 85}, 044903 (2012).




\end{references}
\end{document}